\documentclass[twocolumn,showpacs,preprintnumbers,amsmath,amssymb]{revtex4}
\usepackage{graphicx}

\begin{document}
\draft
\title{Visualization of Coherent Destruction of Tunneling in an Optical Double Well System}
\normalsize
\author{G. Della Valle, M. Ornigotti, E. Cianci$^\dag$, V. Foglietti$^\dag$, P. Laporta, and S. Longhi}
\address{Dipartimento di Fisica and Istituto di Fotonica e Nanotecnologie del CNR,
Politecnico di Milano, Piazza L. da Vinci 32,  I-20133 Milan,
Italy\\
$^\dag$Istituto di Fotonica e Nanotecnologie del CNR,
sezione di Roma, Via Cineto Romano 42, 00156 Roma, Italy}


%
\bigskip
\begin{abstract}
\noindent We report on a direct visualization of coherent
destruction of tunneling (CDT) of light waves in a double well
system which provides an optical analog of quantum CDT as
originally proposed by Grossmann, Dittrich, Jung, and H\"{a}nggi
[Phys. Rev. Lett. {\bf 67}, 516 (1991)]. The driven double well,
 realized by two periodically-curved waveguides
in an Er:Yb-doped glass, is designed so that spatial light
propagation exactly mimics the coherent space-time dynamics of
matter waves in a driven double-well potential governed by the
Schr\"{o}dinger equation. The fluorescence of Er ions is exploited
to image the spatial evolution of light in the two wells, clearly
demonstrating suppression of light tunneling for special ratios
between frequency and amplitude of the driving field.
\end{abstract}

\pacs{42.50.Hz, 03.65.Xp, 42.82.Et}

\maketitle

\newpage
Control of quantum tunneling by external driving fields is a
subject of major relevance in different areas of physics
\cite{Grifoni98,Kohler05}. The driven double-well potential has
provided since more than one decade a paradigmatic model to
investigate tunneling control in such diverse physical systems as
cold atoms in optical traps, superconducting quantum interference
devices, multi-quantum dots and spin systems. Depending on the
strength and frequency of the driving field, suppression
\cite{Grossmann91,Grossmann92} or enhancement \cite{Lin90} of
 tunneling can be achieved. Tunneling enhancement is
usually observed for high field strengths and driving frequencies
close to the classical oscillation frequency at the bottom of each
well. Since the enhancement generally involves a transition
through an intermediate state which is chaotic for strong enough
driving amplitudes, it is often referred to as "chaos-assisted
tunneling" \cite{Grifoni98,Utermann94}. Observations of
chaos-assisted tunneling have been reported in atom optics
experiments \cite{Hensinger01} and in electromagnetic analogs of
quantum mechanical tunneling \cite{Dembowski00,Vorobeichik03}. In
particular, tunneling enhancement has been observed in two coupled
optical waveguides \cite{Vorobeichik03}. In the opposite limit,
Grossmann, H\"{a}nggi and coworkers \cite{Grossmann91} found that,
for certain parameter ratios between amplitude and frequency of
the driving, tunneling can be brought to a standstill. They termed
this effect "coherent destruction of tunneling" (CDT) and, since
then, it has been of continuing interest. Driven tunneling is
related to the problem of periodic nonadiabatic level crossing and
Landau-Zener (LZ) transitions \cite{Grifoni98,Kayanuma94}. In
particular, in the strong modulation limit CDT may be viewed as a
destructive interference effect \cite{Kayanuma94}. In spite of the
great amount of theoretical work devoted to CDT, to date most of
experimental evidences of CDT are rather indirect. In
condensed-matter systems, dephasing and many-particle effects make
tunneling control more involved \cite{Thon04}. In
Ref.\cite{Nakamura01} coherent control of Rabi oscillations in
Josephson-junction circuits irradiated by microwaves has been
reported, however the condition for CDT was not reached. Quantum
interference effects and evidences of CDT in qubit systems have
been recently reported in \cite{Oliver05,Hakonen06}, whereas
suppression of quantum diffusion, also known as dynamic
localization, has been experimentally demonstrated in Refs.
\cite{Keay95,Madison98,Longhi06}. However, CDT is a rather
distinct effect than dynamic localization (see
\cite{Grifoni98,Raghavan96}). For a cleancut demonstration of CDT,
a direct visualization of the dynamics is desirable, which was not
accomplished in all these previous experiments. Engineered
 optical structures, on the other hand, have been recently
demonstrated to provide a very appealing laboratory tool for a
direct visualization of optical analogs of quantum mechanical
phenomena which require a high degree of coherence \cite{Trompeter06}.\\
In this Letter we report on the first visualization of CDT
dynamics using an optical analog of a driven bistable Hamiltonian
based on two tunneling-coupled curved waveguides \cite{Longhi05}
which enables an experimental access to the full space-time
evolution of the corresponding quantum mechanical problem
\cite{Grossmann91}. The structure designed to visualize CDT
consists of a set of two $L=24$-mm-long parallel channel
waveguides, placed at a distance $a=11 \; \mu$m, whose axis is
sinusoidally bent along the propagation distance $z$ with a
bending profile $x_0(z)=A \cos(2 \pi z/ \Lambda)$ [see Fig.1(a)].
The waveguides have been manufactured by the ion-exchange
technique \cite{DellaValle06} in an active Er-Yb phosphate
substrate and probed at $\lambda \simeq 980$ nm wavelength using a
fiber-coupled semiconductor laser [Fig.1(b)] with $\simeq 8 \;
\mu$m mode diameter. A transverse scan of the fiber along the
sample is used to preferentially excite either one of the two
wells. The probing light is partially absorbed by the Yb$^{3+}$
ions (absorption length $\sim 6$ mm), yielding a green
upconversion luminescence arising from the radiative decay of
higher-lying energy levels of Er$^{3+}$ ions \cite{Chiodo06}. By
recording, at successive propagation lengths, the fluorescence
from the top of the sample using a CCD camera connected to a
microscope (magnification factor $\sim 12$) mounted on a
PC-controlled micropositioning system, we could trace with
accuracy the flow of light along the sample.
\begin{figure}
\includegraphics[scale=0.5]{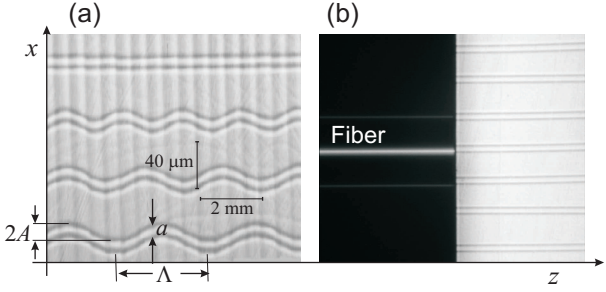}
\caption{Microscope images (top view) of the sample showing (a) a
few set of manufactured coupled optical waveguides, and (b) the
fiber coupling geometry for waveguide excitation.}
\end{figure}
It was previously shown \cite{Longhi05} that the evolution of
light waves in the optical double-well system is formally
equivalent to the dynamics of a periodically-driven
nonrelativistic quantum particle in a double-well potential. In
the scalar and paraxial approximations, light propagation is
described by the equation \cite{Vorobeichik03,Longhi05}
\begin{equation} i \lambdabar \frac{\partial
\psi}{\partial z} = -\frac{\lambdabar^2}{2n_s} \nabla^{2}_{x,y}
\psi + V(x-x_0(z),y) \psi.
\end{equation}
where: $\lambdabar \equiv \lambda / (2 \pi) =1/k$ is the reduced
wavelength, $V(x,y)=[n_{s}^2-n^2(x,y)]/(2 n_s) \simeq n_s-n(x,y)$
is the double-well potential, $n(x,y)$ is the refractive index
profile of the two-waveguide system, and $n_s$ is the reference
(substrate) refractive index. The quantum-optical analogy can be
retrieved after a Kramers-Henneberger transformation
\cite{Longhi05,Longhi06} $x'=x-x_0(z)$, $y'=y$, $z'=z$, $
\phi(x',y',z')=\psi(x',y',z') \exp \left[-i (n_s/ \lambdabar)
\dot{x}_{0}(z') x'
   -i (n_s/ 2 \lambdabar) \int_{0}^{z'} d \xi \; \dot{x }_{0}^2(\xi)
 \right]$, where the dot indicates the derivative with respect to
 $z'$, and after elimination of the $y'$-dependence of the field $\phi$ using a
standard effective index method \cite{Chiang86}. Equation (1) is
then transformed into the following Schr\"{o}dinger
 equation for a particle of mass $n_s$ in the double-well potential
 $V_{e}(x')\simeq n_s-n_{e}(x')$ under the action of a sinusoidal force
 $F(z')$ \cite{Longhi05}:
\begin{equation}
i \lambdabar \frac{\partial \phi}{\partial z'} =
-\frac{\lambdabar^2}{2n_s} \frac{\partial^2 \phi}{\partial x'^2} +
V_{e}(x') \phi-Fx' \phi \equiv \mathcal{H}_0 \phi -Fx' \phi,
 \end{equation}
 \begin{figure}
\includegraphics[scale=0.42]{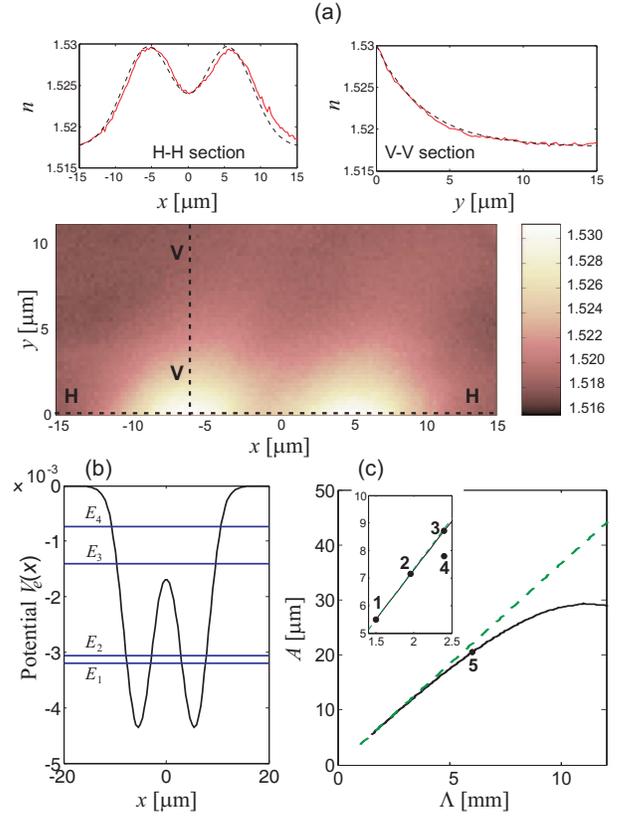} \caption{ (color online)
(a) Measured refractive index profile of the double waveguide
system. (b) Computed profile of the effective 1D double-well
potential. (c) Manifold of quasi-energy crossing in the $(\Lambda,
A)$ plane (solid line). The dashed line has been evaluated from
the first zero of the Bessel function using Eq.(3). The inset is
an enlargement of the linear portion of the manifold,
corresponding to CDT. The five points in the figure correspond to
the geometrical parameters of the optical waveguides manufactured
in our experiment.}
\end{figure}
\noindent where $n_{e}(x')$ is the effective index profile of the
waveguide system and $F(z')=-n_s \ddot{x}_0(z')=(4 \pi^2 A n_s /
\Lambda^2) \cos(2 \pi z'/ \Lambda) $ is the ac force. Note that,
in the optical analog, the Planck constant is played by the
reduced wavelength $\lambdabar$ of photons, whereas the temporal
variable of the quantum problem is mapped into the spatial
propagation coordinate $z'$. CDT is thus simply observed as a
suppression of photon tunneling between
the two waveguides along the propagation direction.\\
 The refractive index profile $n(x,y)$ has been measured by a
refracted-near-field profilometer (Rinck Elektronik) at 670 nm.
The measured 2D index profile is depicted in Fig.2(a), together
with its section profiles along the horizontal ($HH$) and vertical
($VV$) lines. Figure 2(b) shows the corresponding symmetric
double-well potential $V_e(x)\simeq n_s-n_e(x)$ obtained by the
effective index approximation. To derive $n_e(x)$, the measured 2D
index profile $n(x,y)$ was fitted [see dashed curves in Fig.2(a)]
by the relation \cite{Sharma92}
 $n(x,y) \simeq n_s+\Delta n [g(x-a/2)+g(x+a/2)]f(y)$, where
$\Delta n \simeq 0.0124$ is peak index change, $g(x)=[{\rm
erf}((x+w)/D_x)-{\rm} {\rm erf}((x-w)/D_x)]/[2 {\rm erf}(w/D_x)]$
and $f(y)=\exp(-y/D_y)$ define the shape of the index profile
parallel to the surface of the waveguide ($x$-direction) and
perpendicular to the surface ($y$-direction), respectively, $2w
\simeq 5 \; \mu$m is the channel width and $D_x \simeq 4.3 \;
\mu$m, $D_y \simeq 3.3 \; \mu$m  are the lateral and in-depth
diffusion lengths. The numerical computation of the eigenvalues
for the Hamiltonian $\mathcal{H}_0=-\lambdabar^2/(2n_s)
\partial^{2}_{x'}+V_e(x')$ in absence of the driving
force indicates that the double-well potential $V_e$ supports four
bound modes $\xi_l(x')$ whose energies $E_l$
 ($l=1,2,3,4$) are depicted in Fig.2(b) by the four
horizontal solid lines. The linear combinations
$u_{R,L}(x')=[\xi_1(x')\pm \xi_2(x')]/ \sqrt 2$ of the
eigenfunctions $\xi_1(x')$ and $\xi_2(x')$ associated to the
quasi-degenerate energy levels $E_1$ and $E_2$ below the barrier
correspond to photon localization in the right (R) or in left (L)
well of the potential, so that an initial excitation of one of the
two wells, obtained by launching the light into either one of the
two waveguides, is given approximately by the superposition of the
 two lowest eigenstates $\xi_1(x')$ and $\xi_2(x')$. For the straight
 waveguides, the field evolution is dominated by the splitting of
 this doublet, leading to a periodic tunneling of photons
 between the two waveguides with a spatial period $d_{12}=2 \pi \lambdabar
 /(E_2-E_1) \simeq 7.94$ mm. This is clearly shown in Fig.3(a),
 where the measured fluorescence pattern corresponding to
 initial excitation of one of the two waveguides is reported.
For the sake of clearness, in the picture the luminosity level of
the fluorescence, which decreases with propagation distance due to
light absorption, has been gradually rescaled at successive
frames. The measured pattern is very well reproduced by a direct
numerical simulation of Eq.(1),
 performed with a standard beam propagation software (BeamPROP
 4.0). Higher-order bound modes of the double-well potential shown in Fig.2(b)
 can be excited by different beam launching conditions. For instance,
 if the fiber is positioned in the middle between the two
 waveguides, an excitation of the modes $\xi_1(x)$ and $\xi_3(x)$ with
even symmetry is preferentially attained. In this case, the field
evolution is dominated by the splitting of
 the states $\xi_1(x)$ and $\xi_3(x)$, leading to a periodic
 fluorescence pattern with the short spatial period $d_{13}=2 \pi \lambdabar
 /(E_3-E_1) \simeq 545 \; \mu$m [see Fig.3(b)].
\begin{figure}
\includegraphics[scale=0.42]{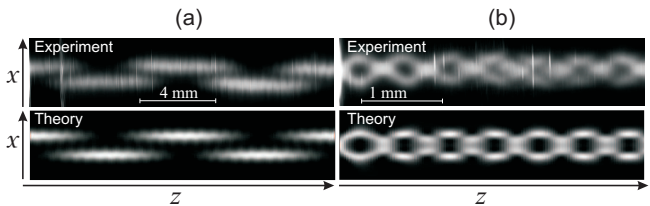} \caption{
Fluorescence light distribution (top view) in the straight
waveguide coupler, and corresponding light intensity distribution
predicted by Eq.(1), for different coupling conditions. (a)
Excitation of one of the two waveguides; (b) Beam launching in the
middle between the two waveguides.}
\end{figure}
The tunneling dynamics in presence of the external force $F$
strongly depends on the amplitude and (spatial) frequency
$\omega=2 \pi / \Lambda$ of the force as compared to the energy
level spacing of the double-well system \cite{Grifoni98}. For
instance, at modulation frequencies comparable with the frequency
spacing $(E_3-E_2)/ \lambdabar$, tunneling is expected to be
enhanced. This case was previously demonstrated for a
two-waveguide optical system in Ref.\cite{Vorobeichik03}.
Conversely, CDT occurs approximately for a modulation frequency in
the range $ (E_2-E_1) \lesssim \lambdabar \omega \lesssim
(E_3-E_2)$ and for a modulation amplitude which corresponds to
exact crossing between the quasienergies $\epsilon_1$ and
$\epsilon_2$ associated to the lowest tunnel doublet
\cite{Grifoni98}. In Fig.2(c) the solid curve shows the manifold
associated to the exact crossing $\epsilon_2=\epsilon_1$ in the
$(\Lambda,A)$ plane, as numerically computed by means of a
two-level approximation of the related driven tunneling problem
\cite{Grifoni98,Grossmann92}. In the high-frequency limit, i.e.
for $\lambdabar \omega \gg (E_2-E_1)$ but $\lambdabar \omega <
(E_3-E_2)$ to avoid the participation in the dynamics of the third
level of energy $E_3$, an approximate expression for the
quasienergy difference $\Delta \epsilon=\epsilon_2-\epsilon_1$
reads \cite{Grifoni98,Grossmann92,Longhi05}
\begin{equation}
\Delta \epsilon=(E_2-E_1)J_0 \left( \frac{2 \pi \mu_{12}
n_s A}{\lambdabar \Lambda} \right)
\end{equation}
\begin{figure}
\includegraphics[scale=0.45]{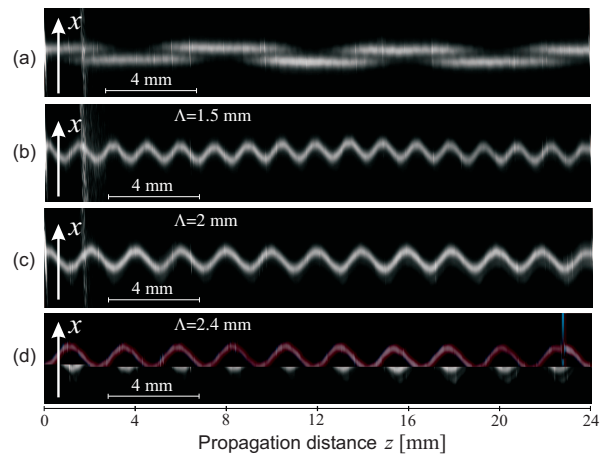} \caption{
Measured fluorescence light distribution in the "waveguide"
reference frame $(x,z)$ for the straight waveguide coupler (a),
and for the three curved waveguide couplers [(b), (c) and (d)]
with increasing period $\Lambda$ and amplitude $A$ corresponding
to points 1,2 and 3 of Fig.2(c).}
\end{figure}
\begin{figure}
\includegraphics[scale=0.47]{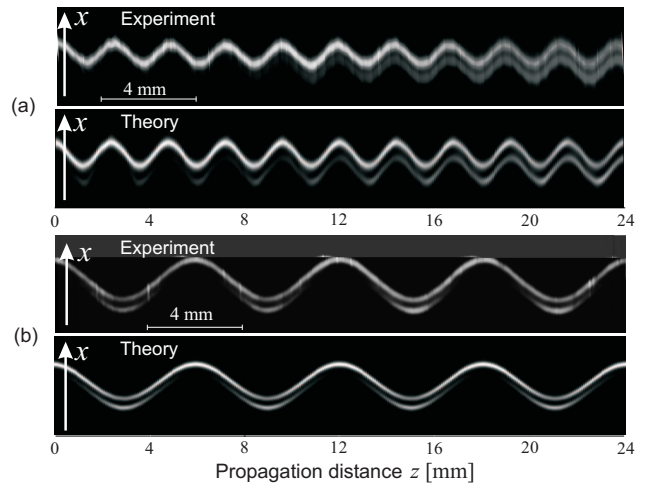} \caption{
Measured fluorescence light pattern, and corresponding photon
density $|\psi(x,0,z)|^2$ predicted by Eq.(1), in curved waveguide
couplers corresponding to (a) break-up of quasienergy crossing
[point 4 in Fig.2(c)], and (b) stroboscopic suppression of
tunneling [point 5 in Fig.2(c)].}
\end{figure}
where $\mu_{12}=\langle \xi_1|x| \xi_2 \rangle$. Considering the first zero of the
Bessel function, the condition $\Delta \epsilon=0$
is thus represented by a straight line in the $(\Lambda,A)$ plane,
which is depicted by the dashed curve in Fig.2(c). Such a curve, however, deviates from the
 exact one as $\Lambda$ increases and approaches $4 \pi \lambdabar /(E_2-E_1) \simeq 16$ mm, where
 the solid curve drops toward zero.
Crossing of the quasienergies is a necessary -but not a
sufficient- condition for the occurrence of CDT. In fact, CDT
requires additionally that the degenerate Floquet states at exact
energy crossing do not show appreciable amplitude oscillations in
one period. A detailed numerical analysis of Eq.(1) shows  that
CDT indeed occurs along the linear portion of the manifold of
Fig.2(c), i.e. for $ \Lambda \lesssim 2.5$ mm, which is
represented by the enlarged inset in the figure. We experimentally
demonstrated the onset of CDT in this portion of the manifold by
manufacturing three curved waveguide couplers corresponding to the
points 1,2 and 3 of Fig.2(c), i.e. to $\Lambda=1.5$ mm, $A \simeq
5.5 \; {\mu}$m (point 1), $\Lambda=2$ mm, $A \simeq 7.3 \; {\mu}$m
(point 2) and $\Lambda=2.4$ mm, $A \simeq 8.8 \; {\mu}$m (point
3). Figure 4 shows the measured fluorescence patterns, as recorded
on the CCD camera from the top of the sample, for the straight
waveguide coupler and for the three curved waveguide couplers when
one of the two waveguides is excited at the input plane. Note
that,  since the fluorescence is proportional to the local photon
density, the patterns in the figure map the profile of $|\psi|^2$
in the "waveguide" reference frame. Therefore, in Figs. 4(b), (c)
and (d) the condition for CDT is clearly demonstrated because the
photon density follows the sinusoidal bending profile $x_0(z)$ of
the initially excited well, without tunneling into the adjacent
waveguide. We also experimentally checked that the observation of
CDT requires that the following two conditions must be {\it
simultaneously} satisfied: (i) quasienergy crossing, and (ii)
absence of appreciable amplitude oscillations of the degenerate
Floquet eigenstates within one oscillation period
\cite{Grifoni98,Grossmann92}. As an example, in Fig.5(a) we show
the measured fluorescence pattern -and corresponding photon
density pattern predicted by the theory- for the curved waveguide
coupler corresponding to point 4 of Fig.2(c) ($\Lambda=2.5$ mm and
$A=7.9 \, \mu$m), in which the condition (i) is not satisfied.
Note that in this case the pattern periodicity is broken and
tunneling is not suppressed, though the tunneling rate is reduced
as compared to the straight waveguide coupler case [compare
Fig.4(a) and Fig.5(a)]. Figure 5(b) shows the measured
fluorescence pattern for the curved waveguide coupler
corresponding to point 5 of Fig.2(c) ($\Lambda=6$ mm and $A \simeq
20.5 \; \mu$m). In this case the condition (i) for quasienergy
crossing is fulfilled, however over one oscillation period the
Floquet eigenstates show non-negligible amplitude oscillations.
Though tunneling is suppressed at the stroboscopic distances
$z=\Lambda , 2 \Lambda, ...$, over one oscillation period an
appreciable fraction of light is observed to tunnel forth and back
between the two waveguides. Such a stroboscopic destruction of
tunneling can be viewed as a result of destructive interference
between successive LZ transitions taking place at periodic level
crossings \cite{Grifoni98,Kayanuma94}, i.e. at the positions
$z=\Lambda/4, 3 \Lambda/4,...$ where $\ddot{x}_0=0$.
This periodic regime, however, does not correspond to a true CDT \cite{Grossmann92}.\\
In conclusion, we reported on the first visualization of photonic
CDT in an optical double-well system which mimics the
corresponding quantum-mechanical problem originally proposed in
\cite{Grossmann91}. The two basic conditions for the observation
of CDT, namely quasienergy crossing and absence of amplitude
oscillations of the degenerate Floquet doublet, have been
experimentally demonstrated.

\end{document}